\documentclass{article}
\usepackage[centertags]{amsmath}
\DeclareMathOperator{\SN}{sn} \DeclareMathOperator{\CN}{cn}
\begin{document}

\title{The classification of all single
travelling wave solutions to Calogero-Degasperis-Focas equation }

\author{ Chengshi Liu \\Department of Mathematics\\Daqing Petroleum Institute\\Daqing 163318, China
\\Email: chengshiliu-68@126.com}

 \maketitle

\begin{abstract}
Under the travelling wave transformation,
Calogero-Degasperis-Focas equation was reduced to an ordinary
differential equation. Using a symmetry group of one-parameter,
this ODE was reduced to a second order linear inhomogeneous ODE.
Furthermore, we applied the change of the variable and complete
discrimination system for polynomial to solve the corresponding
integrals and obtained the classification of all single travelling
wave solutions to Calogero-Degasperis-Focas equation.


\end{abstract}

\section{Introduction}
In the past decade a lot of expansion methods have been proposed
to seek for the travelling wave solutions to nonlinear partial
differential equations(for examples see [1-18]). These methods are
only indirect methods based on the some assumptions about the
forms of the solutions of the equations considered. Applying those
indirect methods, we can't give all single travelling wave
solutions to the equations considered. On the other hand, some
superficially different solutions are essentially the same
solution([19]). So it is worthwhile to give the classifications of
all single travelling wave solutions to those equations. However,
direct integral method is a rather simple and powerful
methods([20-23]),  if a nonlinear equation can be directly reduced
to the integral form as follows:
\begin{equation}
\pm(\xi-\xi_0)=\int \frac{\mathrm{d}u}{\sqrt{p_n(u)}},
\end{equation}
where $p_n(u)$ is an n-th order polynomial, we can give the
classification([20,23]) of all solutions to the right integral in
the Eq.(1) using complete discrimination system for the n-th order
polynomial([20-24]).

But there are some nonlinear differential equations whose reduced
ODE's are the more complex equations.
 Therefore we need more technical methods to get the corresponding reduced ODE and its solutions.
  In the present paper, using the trick of symmetry
group to reduce the Calogero-Degasperis-Focas equation([25]) to an
integrable ODE, and furthermore using the change of variable and
complete discrimination system for polynomial  to solve the
corresponding integral, we obtain the classification of all single
travelling wave solutions to Calogero-Degasperis-Focas equation.
The method and trick used here have general meaning for the
studies of travelling wave solutions to some nonlinear partial
differential equations.

\section{Classification of all single travelling wave solutions to Calogero-Degasperis-Focas equation}

Calogero-Degasperis-Focas equation reads([25])
\begin{equation}
u_t=u_{xxx}-\frac{1}{8}u_{x}^3+(a\exp u+b\exp(-u)+c)u_x.
\end{equation}
Take the travelling wave transformation  $u=u(\xi), \ \
\xi=kx+\omega t $. The corresponding  reduced ODE is as follows:
\begin{equation}
\omega u'=k^3u'''-\frac{k^3}{8}(u')^3+(a\exp u+b\exp(-u)+c)ku'.
\end{equation}
Since the Eq.(2) is invariant under the following one-parameter
group
\begin{eqnarray}
u*=u,\\
 \xi*=\xi+\varepsilon,
\end{eqnarray}
so we can reduce the order of the Eq.(3) by one. For the purpose,
we take $ z=u'$, then we have
$u''=z\frac{\mathrm{d}z}{\mathrm{d}u}$ and
$u'''=z(\frac{\mathrm{d}z}{\mathrm{d}u})^2+z^2\frac{\mathrm{d^2}z}{\mathrm{d}u^2}$.
Instituting those terms into the Eq.(3) and furthermore  letting
$W=z^2$, we have
\begin{equation}
W''-\frac{1}{4}W=\frac{2\omega}{k^3}-\frac{2}{k^2}(a\exp
u+b\exp(-u)+c).
\end{equation}
Using the method of the variation of constants, the general
solution of the Eq.(6) is as follows:
\begin{equation}
W(u)=c_1\exp(\frac{1}{2}u)+c_2\exp(-\frac{1}{2}u)-
\frac{8a}{3k^2}\exp(u)-\frac{8b}{3k^2}\exp(-u)-8(\frac{\omega}{k^3}-\frac{c}{k^2}),
\end{equation}
where $c_1$ and $c_2$ are two arbitrary constants. Hence we have
the following result:

\textbf{Theorem}: the general solution of the Eq.(3) is as
follows:
\begin{eqnarray}\label{IN}
\pm(\xi-\xi_0)=\cr\int\frac{\mathrm{d}u}{\sqrt{c_1\exp(\frac{1}{2}u)+c_2\exp(-\frac{1}{2}u)-
\frac{8a}{3k^2}\exp(u)-\frac{8b}{3k^2}\exp(-u)-8(\frac{\omega}{k^3}-\frac{c}{k^2})}}.
\end{eqnarray}
In order to solve the integral (\ref{IN}),we take a variable
transformation $v=\exp(\frac{1}{2}u)$, the corresponding integral
becomes
\begin{equation}\label{IN2}
\pm\frac{1}{2}(\xi-\xi_0)=\int\frac{\mathrm{d}v}
{\sqrt{a_4v^4+a_3v^3+a_2v^2+a_1v+a_0}}.
\end{equation}
where
\begin{equation}
a_4=-\frac{8a}{3k^2}, \ \ a_3=c_1, \ \
a_2=-\frac{8}{k^2}(\frac{\omega}{k}-c), \ \ a_1=c_2,  \ \
a_0=-\frac{8b}{3k^2}.
\end{equation}
 If $ a<0, $ we take the change of variable
as follows:
\begin{equation}
w=a_4^{\frac{1}{4}}(v+\frac{a_3}{4a_4}),
\end{equation}
Then Eq.(9) becomes
\begin{equation}
\pm\frac{1}{2}a_4^{\frac{1}{4}}(\xi-\xi_0)=\int\frac{\mathrm{d}w}
{\sqrt{w^4+pw^2+qu+r}},
\end{equation}
where
\begin{equation}
p=\frac{a_2}{\sqrt{a_4}},\ \
q=a_4^{-\frac{1}{4}}(\frac{a_3^3}{8a_4^2}-\frac{a_2a_3}{2a_4}+a_1),
\ \
r=a_0-\frac{a_1a_3}{4a_4}+\frac{a_2a_3^2}{16a_4^2}-\frac{3a_3^4}{256a_4^3}.
\end{equation}\\

If $ a>0, $ we take the change of variable as follows:
\begin{equation}
w=(-a_4)^{\frac{1}{4}}(v+\frac{a_3}{4a_4}),
\end{equation}
Then Eq.(9) becomes
\begin{equation}
\pm\frac{1}{2}(-a_4)^{\frac{1}{4}}(\xi-\xi_0)=\int\frac{\mathrm{d}w}
{\sqrt{-(w^4+pw^2+qu+r)}},
\end{equation}
where
\begin{equation}
p=-\frac{a_2}{\sqrt{-a_4}},\ \
q=(-a_4)^{-\frac{1}{4}}(-\frac{a_3^3}{8a_4^2}+\frac{a_2a_3}{2a_4}-a_1),
\ \
r=-a_0-\frac{a_1a_3}{4a_4}-\frac{a_2a_3^2}{16a_4^2}+\frac{3a_3^4}{256a_4^3}.
\end{equation}\\

According to the classification([23]) of the solutions to the ODE
$ (w')^2=\epsilon(w^4+pw^2+qw+r)$, where $\epsilon=\pm1$, we can
give the classification of all single travelling wave solutions to
Calogero-Degasperis-Focas equation. We denote
\begin{equation}
f(w)=w^4+pw^2+qw+r,
\end{equation}\\
and write its complete discrimination system as follows([23]):
\begin{equation*}
\begin{split}
D_1=4, D_2=-p, D_3=8rp-2p^3-9q^2,\\
D_4=4p^4r-p^3q^2+36prq^2-32r^2p^2
-\frac{27}{4}q^4+64r^3,\\
\end{split}
\end{equation*}
\begin{equation}
F_2=9p^2-32pr.
\end{equation}
According to the complete discrimination system (18) and the above
theorem, we obtain the classification of all single travelling
wave solutions to Calogero-Degasperis-Focas equation.\\

Case 1: If $ D_4=0, D_3=0, D_2<0 $, then we have
\begin{equation}
f(w)=((w-l_1)^2+s_1^2)^2,
\end{equation}
where $ l_1, s_1 $ are real numbers, $ s_1>0 $. When $ \epsilon=+1
$, we have
\begin{eqnarray}
u=2\ln\{a_4^{\frac{1}{4}}[s_1\tan
(\frac{1}{2}a_4^{\frac{1}{4}}s_1(\xi-\xi_0))+l_1]-\frac{a_3}{4a_4}\}.
\end{eqnarray}\\

Case 2: If $ D_4=0, D_3=0, D_2=0 $, then we have
\begin{equation}
f(w)=w^4.
\end{equation}
When $ \epsilon=1 $, we have
\begin{eqnarray}
u=2\ln[\pm\frac{2a_4^{-\frac{1}{2}}}{\xi-\xi_0}-\frac{a_3}{4a_4}].
\end{eqnarray}\\

Case 3: If $ D_4=0, D_3=0, D_2>0, E_2>0 $, then we have
\begin{equation}
f(w)=(w-\alpha)^2(w-\beta)^2,
\end{equation}
where $ \alpha, \beta,$ are real numbers, $ \alpha >\beta$. If $
\epsilon=1 $, when $ w>\alpha $ or $ w<\beta, $ we have
\begin{eqnarray}
u=2\ln\{a_4^{-\frac{1}{4}}[\frac{\beta-\alpha}{2}(\coth\frac{\alpha-\beta}
{4}a_4^{\frac{1}{4}}(\xi-\xi_0)-1)+\beta]-\frac{a_3}{4a_4}\};
\end{eqnarray}
when $ \alpha>w>\beta, $ we have
\begin{eqnarray}
u=2\ln\{a_4^{-\frac{1}{4}}[\frac{\beta-\alpha}{2}(\tanh\frac{\alpha-\beta}
{4}a_4^{\frac{1}{4}}(\xi-\xi_0)-1)+\beta]-\frac{a_3}{4a_4}\}.
\end{eqnarray}\\

Case 4: If $ D_4=0, D_3>0, D_2>0 $, then we have
\begin{equation}
f(w)=(w-\alpha)^2(w-\beta)(w-\gamma),
\end{equation}
where $ \alpha, \beta, \gamma $ are real numbers, $ \beta>\gamma.$
If $ \epsilon=1$, when $ \alpha>\beta $ and $ w>\beta, $ or when $
\alpha<\gamma$ and $ w<\gamma, $ we have
\begin{eqnarray}
\exp\{\pm\frac{a_4^{\frac{1}{4}}\sqrt{(\alpha-\beta)(\alpha-\gamma)}(\xi-\xi_0)}{2}\}=\cr\frac{\{\sqrt{[a_4^{\frac{1}{4}}(\exp(\frac{u}{2})+\frac{a_3}{4a_4})-\beta](\alpha-\gamma)}-
\sqrt{(\alpha-\beta)[a_4^{\frac{1}{4}}(\exp(\frac{u}{2})+\frac{a_3}{4a_4})-\gamma]}
\}^2}{|a_4^{\frac{1}{4}}(\exp(\frac{u}{2})+\frac{a_3}{4a_4})-\alpha|};
\end{eqnarray}
when $ \alpha>\beta $ and $ w<\gamma, $ or when $ \alpha<\gamma$
and $ w<\beta, $ we have
\begin{eqnarray}
\exp\{\pm\frac{a_4^{\frac{1}{4}}\sqrt{(\alpha-\beta)(\alpha-\gamma)}(\xi-\xi_0)}{2}\}=\cr\frac{\{\sqrt{[a_4^{\frac{1}{4}}
(\exp(\frac{u}{2})+\frac{a_3}{4a_4})-\beta](\gamma-\alpha)}-\sqrt{(\beta-\alpha)[a_4^{\frac{1}{4}}(\exp(\frac{u}{2})+\frac{a_3}{4a_4})
-\gamma]}\}^2}{|a_4^{\frac{1}{4}}(\exp(\frac{u}{2})+\frac{a_3}{4a_4})-\alpha|};
\end{eqnarray}
when $ \beta>\alpha>\gamma, $ we have
\begin{eqnarray}
\pm\sin\{\frac{a_4^{\frac{1}{4}}\sqrt{(\beta-\alpha)(\alpha-\gamma)}(\xi-\xi_0)}{2}\}=\cr\frac{[a_4^{\frac{1}{4}}(\exp(\frac{u}{2})+\frac{a_3}{4a_4})-\beta](\alpha-\gamma)+(\alpha-\beta)
[a_4^{\frac{1}{4}}(\exp(\frac{u}{2})+\frac{a_3}{4a_4})-\gamma]}{|[a_4^{\frac{1}{4}}(\exp(\frac{u}{2})+\frac{a_3}{4a_4})-\alpha](\beta-\gamma)|}.
\end{eqnarray}\\

If $ \epsilon=-1, $ when $ \alpha>\beta $ and $ w>\beta, $ or when
$ \alpha<\gamma$ and $ w<\gamma, $ we have
\begin{eqnarray}
\exp\{\pm\frac{(-a_4)^{\frac{1}{4}}\sqrt{(\alpha-\beta)(\alpha-\gamma)}(\xi-\xi_0)}{2}\}=\cr\frac{\{\sqrt{[(-a_4)^{\frac{1}{4}}(\exp(\frac{u}{2})+\frac{a_3}{4a_4})-\beta](\alpha-\gamma)}-\sqrt{(\alpha-\beta)
[(-a_4)^{\frac{1}{4}}(\exp(\frac{u}{2})+\frac{a_3}{4a_4})-\gamma]}\}^2}{|(-a_4)^{\frac{1}{4}}(\exp(\frac{u}{2})+\frac{a_3}{4a_4})-\alpha|};
\end{eqnarray}
when $ \alpha>\beta $ and $ w<\gamma, $ or When $ \alpha<\gamma$
and $ w<\beta, $ we have
\begin{eqnarray}
\exp\{\pm\frac{(-a_4)^{\frac{1}{4}}\sqrt{(\alpha-\beta)(\alpha-\gamma)}(\xi-\xi_0)}{2}\}=\cr\frac{\{\sqrt{[(-a_4)^{\frac{1}{4}}(\exp(\frac{u}{2})+\frac{a_3}{4a_4})-\beta]
(\gamma-\alpha)}-\sqrt{(\beta-\alpha)[(-a_4)^{\frac{1}{4}}(\exp(\frac{u}{2})+\frac{a_3}{4a_4})-\gamma]}\}^2}{|(-a_4)^{\frac{1}{4}}[\exp(\frac{u}{2})+\frac{a_3}{4a_4}]-\alpha|};
\end{eqnarray}
when $ \beta>\alpha>\gamma, $ we have
\begin{eqnarray}
\pm\sin\{\frac{(-a_4)^{\frac{1}{4}}\sqrt{(\beta-\alpha)(\alpha-\gamma)}(\xi-\xi_0)}{2}\}=\cr{\frac{[(-a_4)^{\frac{1}{4}}(\exp(\frac{u}{2})+\frac{a_3}{4a_4})-\beta]
(\alpha-\gamma)+(\alpha-\beta)[(-a_4)^{\frac{1}{4}}(\exp(\frac{u}{2})+\frac{a_3}{4a_4})-\gamma]}{|[(-a_4)^{\frac{1}{4}}(\exp(\frac{u}{2})+\frac{a_3}{4a_4})-\alpha](\beta-\gamma)|}}.
\end{eqnarray}\\

Case 5: If $ D_4=0, D_3=0, D_2>0, E_2=0 $, then we have
\begin{equation}
f(w)=(w-\alpha)^3(w-\beta),
\end{equation}
where $ \alpha, \beta $ are real numbers. If $ \epsilon=1, $ when
$ w>\alpha, w>\beta; $ or $ w<\alpha, w<\beta, $ We have
\begin{equation}
u=2\ln\{a_4^{-\frac{1}{4}}[\frac{4(\alpha-\beta)}{\frac{1}{4}a_4^{\frac{1}{2}}(\alpha-\beta)^2(\xi-\xi_0)^2-4}+\alpha]-\frac{a_3}{4a_4}\}.
\end{equation}
If $ \epsilon=-1, $ when $ w>\alpha, w<\beta; $ or $ w<\alpha,
w>\beta, $ we have
\begin{equation}
u=2\ln\{(-a_4)^{-\frac{1}{4}}[\frac{4(\beta-\alpha)}{4+\frac{1}{4}(-a_4)^{\frac{1}{2}}(\alpha-\beta)^2(\xi-\xi_0)^2}+\alpha]-\frac{a_3}{4a_4}\}.
\end{equation}\\

Case 6: If $ D_4=0, D_2D_3<0 $, then we have
\begin{equation}
f(w)=(w-\alpha)^2((w-l_1)^2+s_1^2),
\end{equation}
where $ \alpha, l_1 $ and $ s_1 $ are real numbers. If $
\epsilon=1, $ we have
\begin{eqnarray}
u=2\ln\{a_4^{-\frac{1}{4}}\times\cr[\frac{\exp{[\pm\frac{1}{2}a_4^{\frac{1}{4}}\sqrt{(\alpha-l_1)^2+s_1^2}(\xi-\xi_0)]}-\gamma+(2-\gamma)\sqrt{(\alpha-l_1)^2+s_1^2}}
{\{\exp{[\pm\frac{1}{2}a_4^{\frac{1}{4}}\sqrt{(\alpha-l_1)^2+s_1^2}(\xi-\xi_0)]}-\gamma\}^2-1}]\cr-\frac{a_3}{4a_4}\}.
\end{eqnarray}
where
\begin{equation}
 \gamma=\frac{\alpha-2l_1}{\sqrt{(\alpha-l_1)^2+s_1^2}}.
\end{equation}\\

Case 7: If $ D_4>0, D_3>0, D_2>0 $, then we have
\begin{equation}
f(w)=(w-\alpha_1)(w-\alpha_2)(w-\alpha_3)(w-\alpha_4),
\end{equation}
where $ \alpha_1, \alpha_2, \alpha_3, \alpha_4 $ are real numbers.
and $\alpha_1>\alpha_2> \alpha_3>\alpha_4.$ If $\epsilon=1 $, when
$ W>\alpha_1 $ or $ w<\alpha_4, $ we have
\begin{eqnarray}
u=2\ln\{a_4^{-\frac{1}{4}}\times\cr[\frac{\alpha_2(\alpha_1-\alpha_4)\SN^2{(\frac{a_4^{\frac{1}{4}}\sqrt{(\alpha_1-\alpha_3)(\alpha_2-\alpha_4)}}{4}(\xi-\xi_0),m)-\alpha_1(\alpha_2-\alpha_4)}}
{(\alpha_1-\alpha_4)\SN^2{(\frac{a_4^{\frac{1}{4}}\sqrt{(\alpha_1-\alpha_3)(\alpha_2-\alpha_4)}}{4}(\xi-\xi_0),m)-(\alpha_2-\alpha_4)}}]\cr-\frac{a_3}{4a_4}\};
\end{eqnarray}
when $ \alpha_2>w>\alpha_3, $ we have
\begin{eqnarray}
u=2\ln\{a_4^{-\frac{1}{4}}\times\cr[\frac{\alpha_4(\alpha_2-\alpha_3)\SN^2{(\frac{a_4^{\frac{1}{4}}\sqrt{(\alpha_1-\alpha_3)(\alpha_2-\alpha_4)}}{4}(\xi-\xi_0),m)-\alpha_3(\alpha_2-\alpha_4)}}
{(\alpha_2-\alpha_3)\SN^2{(\frac{a_4^{\frac{1}{4}}\sqrt{(\alpha_1-\alpha_3)(\alpha_2-\alpha_4)}}{4}(\xi-\xi_0),m)-(\alpha_2-\alpha_4)}}]\cr-\frac{a_3}{4a_4}\}.
\end{eqnarray}

If $ \epsilon=-1, $ when $ \alpha_1>w>\alpha_2, $ we have
\begin{eqnarray}
u=2\ln\{(-a_4)^{-\frac{1}{4}}\times\cr[\frac{\alpha_3(\alpha_1-\alpha_2)\SN^2{(\frac{(-a_4)^{\frac{1}{4}}\sqrt{(\alpha_1-\alpha_3)(\alpha_2-\alpha_4)}}{4}(\xi-\xi_0),m)-\alpha_2(\alpha_1-\alpha_3)}}
{(\alpha_1-\alpha_2)\SN^2{(\frac{(-a_4)^{\frac{1}{4}}\sqrt{(\alpha_1-\alpha_3)(\alpha_2-\alpha_4)}}{4}(\xi-\xi_0),m)-(\alpha_1-\alpha_3)}}]\cr-\frac{a_3}{4a_4}\};
\end{eqnarray}

when $ \alpha_3>w>\alpha_4, $ we have
\begin{eqnarray}
u=2\ln\{(-a_4)^{-\frac{1}{4}}\times\cr[\frac{\alpha_1(\alpha_3-\alpha_4)\SN^2{(\frac{(-a_4)^{\frac{1}{4}}\sqrt{(\alpha_1-\alpha_3)(\alpha_2-\alpha_4)}}{4}(\xi-\xi_0),m)-\alpha_4(\alpha_3-\alpha_1)}}
{(\alpha_3-\alpha_4)\SN^2{(\frac{(-a_4)^{\frac{1}{4}}\sqrt{(\alpha_1-\alpha_3)(\alpha_2-\alpha_4)}}{4}(\xi-\xi_0),m)-(\alpha_3-\alpha_1)}}]\cr-\frac{a_3}{4a_4}\},
\end{eqnarray}
where $ m^2=\frac{(\alpha_1-\alpha_2)( \alpha_3-
\alpha_4)}{(\alpha_1- \alpha_3)( \alpha_2- \alpha_4)} $.\\

Case 8: If $ D_4<0, D_2D_3\geq0 $, then we have
\begin{equation}
f(w)=(w-\alpha)(w-\beta)((w-l_1)^2+s_1^2),
\end{equation}
where $ \alpha, \beta, l_1 $ and $ s_1 $ are real numbers, and $
\alpha>\beta, s_1>0 $, we have
\begin{eqnarray}
u=2\ln\{(\epsilon a_4)^{-\frac{1}{4}}[\frac{a\CN{(\frac{(\epsilon
a_4)^{\frac{1}{4}}\sqrt{-\epsilon2s_1m_1(\alpha-\beta)}}{4mm_1}(\xi-\xi_0),m)}+b}
{c\CN{(\frac{(\epsilon
a_4)^{\frac{1}{4}}\sqrt{-\epsilon2s_1m_1(\alpha-\beta)}}{4mm_1}(\xi-\xi_0),m)}+d}]-\frac{a_3}{4a_4}\}.
\end{eqnarray}
where
\begin{eqnarray*}
c=\alpha-l_1-\frac{s_1}{m_1}, d=\alpha-l_1-s_1m_1,\\
a=\frac{1}{2}[(\alpha+\beta)c-(\alpha-\beta)d],
b=\frac{1}{2}[(\alpha+\beta)d-(\alpha-\beta)c],
\end{eqnarray*}
\begin{equation}
E=\frac{s_1^2+(\alpha-l_1)(\beta-l_1)}{s_1(\alpha-\beta)}, \ \
m_1=E\pm\sqrt{E^2+1}, m^2=\frac{1}{1+m_1^2},
\end{equation}
we choose $ m_1 $ such that $ \epsilon m_1<0. $\\

Case 9: If $ D_4>0, D_2D_3\leq0, $ then we have
\begin{equation}
f(w)=((w-l_1)^2+s_1^2)(w-l_2)^2+s_2^2),
\end{equation}
where $ l_1, l_2, s_1 $ and $ s_2 $ are real numbers, and $
s_1>s_2>0 $. If $ \epsilon=1, $  we have
\begin{equation}
u=2\ln\{a_4^{-\frac{1}{4}}[\frac{a\SN{(\eta(\xi-\xi_0),m)}+b\CN{(\eta(\xi-\xi_0),m)}}{c\SN{(\eta(\xi-\xi_0),m)}+d\CN{(\eta(\xi-\xi_0),m)}}]-\frac{a_3}{4a_4}\},
\end{equation}
where
\begin{eqnarray*}
c=-s_1-\frac{s_2}{m_1}, \ \ d=l_1-l_2, \ \  a=l_1c+s_1d, \ \ b=l_1d-s_1c,\\
E=\frac{s_1^2+s_2^2+(l_1-l_2)^2}{2s_1s_2}, \ \ m_1=E+\sqrt{E^2-1},
\end{eqnarray*}
\begin{equation}
m^2=1-\frac{1}{m_1^2}, \ \
\eta=\frac{1}{2}a_4^{\frac{1}{4}}s_2\sqrt{\frac{m_1^2c^2+d^2}{c^2+d^2}}.
\end{equation}\\
These are all possible cases, so we obtain the complete
classification of all single travelling wave solutions to
Calogero-Degasperis-Focas equation.

\section{Conclusions}

In summary, although the classifications of all single travelling
wave solutions to nonlinear partial differential equations is a
rather difficult problem. But there are a lot of nonlinear
differential equations whose all travelling wave solutions can be
obtained using direct integral method and complete discrimination
system for polynomial. On the other hand, if a nonlinear
differential equation whose reduced ODE can't be obtained by
simple integral method, then we will need find more powerful
tricks and methods to do this thing. In this paper, we use
symmetry group to reduce the order of ODE, and furthermore reduce
the equation to an integrable ODE. Using complete discrimination
system for polynomial, we obtain the classifications of all single
travelling wave solutions to some nonlinear partial differential
equations.  The methods and tricks used here can be expected to
develop to solve more complex and more extensive equations.

\end{document}